\pdfoutput=1
\documentclass[preprint,12pt]{elsarticle}

\usepackage{amsmath}
\usepackage{amssymb}
\usepackage{amstext}
\usepackage{amsthm}
\usepackage{xfrac}
\usepackage{mathtools}
\usepackage[ruled,vlined]{algorithm2e}
\usepackage{fixmath}

\usepackage{caption}
\usepackage{subfig}
\graphicspath{ {./figs/} }
\DeclareGraphicsExtensions{.pdf}

\usepackage{tikz}
\usepackage{xspace}
\usepackage{xcolor}

\usepackage{lineno}
\usepackage{multirow}
\usepackage{enumerate}
\usepackage{environ}
\usepackage{placeins}
\usepackage{hyperref}
\usepackage{url}

\newcommand{\minimize}[1]{\ensuremath{\underset{#1}{\text{minimize}}}}
\newcommand{\subjectto}{\ensuremath{\text{subject to}}}
\NewEnviron{multiline}{
\begin{equation}\begin{split}
    \BODY
\end{split}\end{equation}
}
\newdefinition{rmk}{Remark}

\newif\iffinal
\finaltrue

\iffinal
  \newcommand{\cs}[1]{}
  \newcommand{\todd}[1]{}
  \newcommand{\alp}[1]{}
\else
  \definecolor{darkgreen}{rgb}{0,0.5,0}
  \newcommand{\cs}[1]{\textcolor{red}{[\textit{CS: #1}]}}
  \newcommand{\todd}[1]{\textcolor{darkgreen}{[\textit{Todd: #1}]}}
  \newcommand{\alp}[1]{\textcolor{blue}{[\textit{Alp: #1}]}}
\fi

\newif\ifanlreport

\journal{Journal of Computational Physics}

\begin{document}

\begin{frontmatter}

\ifanlreport
  \onecolumn
\pagestyle{empty}
  
\vspace{1.75in}

\begin{centering}

ARGONNE NATIONAL LABORATORY

9700 South Cass Avenue

Argonne, Illinois  60439

\vspace{1.5in}

{\large \textbf{Training neural networks under physical constraints using a stochastic augmented Lagrangian approach}}

\vspace{.5in}

\textbf{A. Dener, M.A. Miller, R.M. Churchill, T. Munson, and C.S. Chang}

\vspace{.5in}

Mathematics and Computer Science Division

\vspace{.25in}

Preprint ANL/MCS-P9367-0920

\vspace{.5in}

September 2020

\end{centering}

\vspace{2.0in}

\bigskip

\par\noindent
\footnotetext [1]
{
This work was supported by the U.S. Department of Energy, Office of Science, Office of
Advanced Scientific Computing Research, Scientific Discovery through Advanced
Computing (SciDAC) program via the FASTMath Institute under Contract
No. DE-AC02-06CH11357 at Argonne National Laboratory and via the Partnership Center for High-fidelity Boundary Plasma Simulation at Princeton Plasma Physics Laboratory under the Contract No. DE-AC02–09CH11466.

The XGC simulations used computational resources at the Argonne (Theta) and Oak Ridge (Summit)
Leadership Computing Facilities, DOE Office of Science User Facilities supported under 
Contracts DE-AC02-06CH11357 and DE-AC05-00OR22725, respectively.
}

\newpage

\vspace*{\fill}
\begin{center}
\fbox{
\parbox{4in}{
The submitted manuscript has been created by UChicago Argonne, LLC, Operator of Argonne 
National Laboratory (``Argonne''). Argonne, a U.S. Department of Energy Office of Science 
laboratory, is operated under Contract No. DE-AC02-06CH11357. The U.S. Government retains 
for itself, and others acting on its behalf, a paid-up nonexclusive, irrevocable worldwide 
license in said article to reproduce, prepare derivative works, distribute copies to the 
public, and perform publicly and display publicly, by or on behalf of the Government. 
The Department of Energy will provide public access to these results of federally 
sponsored research in accordance with the DOE Public Access 
Plan. \url{http://energy.gov/downloads/doe-public-accessplan}
}}
\end{center}
\vfill

\newpage
\pagestyle{plain}
\setcounter{page}{1}

\fi 

\title{Training neural networks under physical constraints using a stochastic augmented Lagrangian approach}

\author[ANL]{A. Dener}
\author[PPPL]{M.A. Miller}
\author[PPPL]{R.M. Churchill}
\author[ANL]{T. Munson}
\author[PPPL]{C.S. Chang}

\address[ANL]{Argonne National Laboratory, 9700 S Cass Ave, Lemont, IL 60439, USA}
\address[PPPL]{Princeton Plasma Physics Laboratory, 100 Stellarator Road, Princeton, NJ 08540, USA}

\begin{abstract}
We investigate the physics-constrained training of an encoder-decoder neural network for approximating the Fokker-Planck-Landau collision operator in the 5-dimensional kinetic fusion simulation in XGC. To train this network, we propose a stochastic augmented Lagrangian approach that utilizes pyTorch's native stochastic gradient descent method to solve the inner unconstrained minimization subproblem, paired with a heuristic update for the penalty factor and Lagrange multipliers in the outer augmented Lagrangian loop. Our training results for a single ion species case, with self-collisions and collision against electrons, show that the proposed stochastic augmented Lagrangian approach can achieve higher model prediction accuracy than training with a fixed penalty method for our application problem, with the accuracy high enough for practical applications in kinetic simulations.
\end{abstract}


\end{frontmatter}

\section{Introduction}
\label{sec:intro}
XGC is a massively parallel Lagrangian particle-in-cell based gyrokinetic code, utilizing a 5-dimensional Eulerian mesh for dissipative operations such as Coulomb collisions, optimized for simulating the edge region of fusion devices such as tokamaks~\cite{ku2018}. With the increasing availability of more powerful high performance computing resources, there has been growing interest in using XGC to solve problems involving many plasma species that are relevant to ITER plasma conditions where the tungsten wall makes the plasma to be contaminated by many different charge-state tungsten ion species. Even in the cold edge plasma where the ionization level may not be very high, tungsten ions can easily have $\sim 10$ different ionization species. In the hot core plasma, the large number of ionization species may be bundled into smaller number but still requires minimum of $\sim 10$ bundled species for a reasonable physics understanding.

However, the present Fokker-Planck-Landau (FPL) collision operator in XGC exhibits quadratic scaling with the number of species and presents a computational bottleneck for the study of many species fusion cases. In the present work, we apply machine learning (ML) techniques to address this limitation.

The Fokker-Planck collision operator in the Landau form is given as 
\begin{equation}
    \frac{df_a}{dt} = \sum_b C_{ab}(f_a;f_b) = -\sum_b \frac{e_a^2e_b^2\ln{\Lambda_{ab}}}{8\pi\epsilon_0^2m_a} \nabla_v \cdot \int \textbf{U} \cdot \left(\frac{f_a}{m_b} \nabla_v' f_b' - \frac{f_b'}{m_a} \nabla_v f_a\right)d^3v',
    \label{eq:fpl}
\end{equation}
where $a$ and $b$ subscripts denote separate plasma species, $f_a$ and $f_b$ are the velocity particle distribution functions, $e$ are the charges, $m$ are the masses, $\ln{\Lambda_{ab}}$ is the Coulomb logarithm, and $\textbf{U}$ is a tensor that is a function of the relative velocity vector\cite{landau1936transport}.

XGC employs an implicit Picard iteration scheme to solve \eqref{eq:fpl}, where the changes in the distribution functions are computed as \cite{yoon2014,hager2016}
\begin{multiline}
    \Delta f_a &= \Delta t \cdot \left[ C_{aa}(f_a; f_a) + C_{ab}(f_a; f_b) \right], \\
    \Delta f_b &= \Delta t \cdot \left[ C_{bb}(f_a; f_a) + C_{ba}(f_a; f_b) \right].
\end{multiline}
Here, the $C_{aa}$ and $C_{bb}$ operators describe self-collision while $C_{ab}$ and $C_{ba}$ operators describe collisions between different species. With each additional species in the problem, XGC must compute an additional function distribution perturbation \textit{and} an additional inter-species collision operator for each species perturbation, resulting in a $\mathcal{O}(n^2)$ scaling in computational cost where $n$ is the number of species. This motivates the exploration of alternative solution or approximation approaches that exhibit more favorable scaling to enable the study of many-species cases.

In recent years, there has been increased focus on using ML models to approximate various operators and kernels in the simulation of physical phenomena. Much of this research has been driven by efforts to use deep learning techniques to generate turbulence closures for Reynolds-averaged Navier-Stokes methods, scale-resolving simulations, and large eddy simulations~\cite{tracey2015machine,gamahara2017searching,beck2019deep,zhu2019machine,zhang2019recent,duraisamy2019turbulence}. Similarly, we seek an ML model that can adequately and efficiently approximate the FPL operator in \eqref{eq:fpl}, and integrate into XGC and enable the simulation of fusion problems with many plasma species.

An important consideration is that the ML model must generate operators and kernels that conserve mass, momentum and energy properties in the simulation. Similar considerations of physical consistency with respect to boundary conditions, spatial constraints and conservation laws emerge routinely in scientific ML applications. There have recently been a number of successful efforts to explicitly enforce such constraints to machine precision, either via physics-informed neural networks where the architecture of the network mimics the mathematical structure of the governing equations~\cite{raissi2019physics}, or through the use of fixed constraint projection layers that manipulate the model output~\cite{beucler2019enforcing,mohan2020embedding}. 

These explicit or ``hard'' constraint enforcement techniques, however, are only applicable to simple governing equations or constraints for which there are efficient differential or boundary operators available. For problems with complex governing equations or challenging constraints, researchers have modified the ML training loss function with a penalty term 
to guide the model parameters toward model outputs that simultaneously minimize the prediction error and the constraint violation~\cite{kim2019deep}. This is the approach we have adopted in a previous effort to construct an ML model for the FPL operator in XGC~\cite{miller2020penalty}.

Unfortunately, the fixed penalty method for implicit or ``soft'' enforcement of such constraints face a number of challenges. Although the constrained ML model trained with this approach yields some improvement in conservation properties relative to the unconstrained model, the constraint violation still remains well above the error margins we consider acceptable for integration with the full XGC simulation workflow that employs thousands of collision steps hence requires at least $10^{-5}$ level relative error per single collision operation to limit the time integrated error to within several percent, at least. Furthermore, the fixed penalty method features a scalar penalty weight for balancing the relative magnitudes of the model error and constraint violation terms in the loss function. This parameter requires cumbersome hand-tuning to prevent either the model error or the constraint violation from overwhelming the other while training the model.

In the present work, we propose a stochastic augmented Lagrangian method for solving nonlinearly constrained ML training problems. Augmented Lagrangian methods, originally known as the method of multipliers, were first introduced by Hestenes~\cite{hestenes1969multiplier} and Powell~\cite{powell1969method} for nonlinearly constrained optimization problems. This approach decomposes the constrained problem into a sequence of unconstrained problems where Lagrange multipliers and the penalty parameter are dynamically updated in the outer loop iterations based on the progress of the inner loop optimization problem. Our proposed method can be viewed as a heuristic adaptation of the augmented Lagrangian approach to ML training problems where the unconstrained optimization problem in the inner loop is solved using a conventional stochastic gradient descent (SGD) method. The introduction of the Lagrange multipliers and the automated control of the penalty parameter in this approach promises to address the issues we have experienced with the fixed penalty method.

For the first introduction, we apply the present technique to deuteron ions, that includes the deuteron self-collisions and collision against electrons.   
\[
\Delta f_i = \Delta t \cdot \left[ C_{ii}(f_i; f_i) + C_{ie}(f_i; f_e) \right].
\]
Here, the subscript ``i'' represent deuterons and ``e'' represent electrons.  In the present application, there is no other ion species. Even without explicitly evaluating the collisional change in the electron distribution function, the present application example can still be used for the ``ion temperature gradient'' turbulence simulation in which the electrons simply respond as an adiabatic fluid. The data is from the actual XGC simulations, to which the training and inference are applied, with complete ion and electron FPL collisions. Extension of the application to the ML evaluation of the collisional change in the electron electron distribution function will be the subject of a subsequent report. Generalization to multiple ion species will then follow.

The remainder of the paper is organized as follows. We begin by reviewing our formulation of the physics-constrained supervised training problem for deep neural networks in Section \ref{sec:ml_overview} and introduce the stochastic augmented Lagrangian method for constrained supervised training in Section \ref{sec:stoch_auglag}. We then present the numerical results for the constrained ML model trained with the augmented Lagrangian method in Section \ref{sec:results} and conclude with a review of our observations in Section \ref{sec:conclusions}.

\section{Physics-Constrained Supervised Learning}
\label{sec:ml_overview}
Neural network (NN) models are often constructed via supervised learning using simulation data. This training process is a data-driven optimization problem for finding the model parameters (e.g., NN weights) that minimize a ``loss'' function representing the error between the model prediction and the simulation outputs for a given set of inputs. For instance, the supervised learning problem using the canonical mean squared error (MSE) loss function is
\begin{equation}\label{eq:mse_loss}
    \minimize{p} \quad \mathcal{J}(p) = \frac{1}{2N}\sum_i^N \| \mathcal{M}(p, x_i) - y_i \|_2^2,
\end{equation}
where $p \in \mathbb{R}^n$ are the NN weights, and $N$ is the number of inputs $x_i \in \mathbb{R}^s$ and corresponding simulation outputs $y_i \in \mathbb{R}^t$ that we want to approximate using the NN model $\mathcal{M}:\mathbb{R}^{n \times s} \rightarrow \mathbb{R}^t$.

The stochastic gradient descent (SGD) family of methods~\cite{robbins1951stochastic,kiefer1952stochastic}, which have become the standard for solving NN training problems, are stochastic generalizations of gradient descent methods where the full gradient is replaced by a sub-sampled approximation. For the MSE loss function in \eqref{eq:mse_loss}, the sub-sampled gradient is given as
\begin{equation}\label{eq:stoch_grad}
    \nabla_p \mathcal{J}(p) \approx \frac{1}{|\mathcal{N}|}\sum_{j \in \mathcal{N}} (\mathcal{M}(p, x_j) - y_j)^T \nabla_p \mathcal{M}(p, x_j),
\end{equation}
where $\mathcal{N} = \{ 1 \leq j \leq N \; : \; |\mathcal{N}| \ll N \}$ denotes the indices for a randomly selected mini-batch of the training data. 

In practice, an SGD solution to the training problem \eqref{eq:mse_loss} randomly shuffles the entire training data set, splits it into batches of equal size, and takes a gradient descent direction for each generated batch of data. When there are no more unused batches left, the training data is randomly shuffled again and a new set of unique random batches are generated for SGD to iterate through. This sub-sampling strategy significantly reduces the computational cost of evaluating gradients for large training data sets, and has become an essential component of large-scale machine learning applications.

In the present work, we wish to find the NN weights $p$ that not only minimize the model output error with respect to the FPL collision operator, but also generate outputs that obey conservation laws essential to the XGC simulation workflow. As we will demonstrate later in Section \ref{sec:results}, the unconstrained learning problem in \eqref{eq:mse_loss} cannot construct a model with the desired conservation properties even if the underlying training data is conservative. This suggests that we must recast the original supervised learning problem in \eqref{eq:mse_loss} as an equality-constrained optimization problem,
\begin{multiline}\label{eq:cnstr_learn}
    \minimize{p} &\quad \mathcal{J}(p), \\
    \subjectto &\quad \mathcal{C}(p) = 0,
\end{multiline}
where $\mathcal{C}:\mathbb{R}^n \rightarrow \mathbb{R}^m$ computes how much the model output violates conservation laws for mass, momentum and energy properties.


Nonlinear quality-constrained optimization problems, such as the constrained learning problem in \eqref{eq:cnstr_learn}, are often solved using constrained sequential quadratic programming (SQP) techniques~\cite{nocedal2006numerical}. However, this approach requires the ability to compute accurate gradients. In machine learning applications with large quantities of data, we can only generate stochastic mini-batch gradients that exhibit large inaccuracies. To the best of our knowledge, constrained SQP methods have not yet been generalized with stochastic approximations.

Nonetheless, it is possible to convert these constrained optimization problems to unconstrained problems by adding a constraint violation penalty term to the loss function, such that
\begin{equation}\label{eq:fixed_penalty}
    \minimize{p} \quad \mathcal{J}(p) + \mu \sum_i^m \mathcal{C}^2_i(p).
\end{equation}
This yields the fixed penalty method where the penalty factor $\mu$ determines the balance of priorities in the training between the loss function and the constraint enforcement. The SGD method can be directly applied to \eqref{eq:fixed_penalty} to solve this penalized training problem.

We have previously investigated the fixed penalty method to generate the NN model for approximating the FPL collision operator~\cite{miller2020penalty}. Although the inclusion of the constraints as a penalty in the loss function improved the model's conservation properties, the constraint violation remained too high for the model to adequately replace the collision operator in the XGC simulation. This motivates us to explore alternative approaches to solve \eqref{eq:cnstr_learn}, such as the augmented Lagrangian method we discuss in the next section, that can achieve an average constraint violation in the conservation properties no greater than $10^{-6}$.

\section{Stochastic Augmented Lagrangian}
\label{sec:stoch_auglag}
The augmented Lagrangian method~\cite{hestenes1969multiplier,powell1969method}, which we will call ``aug-Lag'' for the remainder of the paper, decomposes a general equality-constrained optimization problem,
\begin{multiline}\label{eq:general_opt}
    \underset{p}{\text{minimize}} &\quad \mathcal{J}(p), \\
    \text{subject to} &\quad \mathcal{C}(p) = 0,
\end{multiline}
into a sequence of unconstrained optimization problems denoted by the $k$ subscript,
\begin{equation}\label{eq:auglag_inner}
    \underset{p}{\text{minimize}} \quad \mathcal{\hat{L}}(p, \lambda_k, \mu_k),
\end{equation}
using the augmented Lagrangian merit function,
\begin{equation}\label{ref:auglag_merit}
    \mathcal{\hat{L}}(p, \lambda, \mu) = \mathcal{J} + \lambda^T\mathcal{C}(p) + \frac{\mu}{2}\|\mathcal{C}(p)\|_2^2,
\end{equation}
where $p \in \mathbb{R}^n$ are the optimization parameters, $\mathcal{J}:\mathbb{R}^n \rightarrow \mathbb{R}$ is the objective function, $\mathcal{C}:\mathbb{R}^n \rightarrow \mathbb{R}^m$ are the equality constraints, $\lambda \in \mathbb{R}^m$ are Lagrange multipliers, and $\mu$ is the augmented Lagrangian penalty factor. When applied to the training of the ML model for the XGC FPL collision operator, the objective function represents the mean squared error (MSE) loss function introduced in Section \ref{sec:ml_overview} with neural network (NN) weights as optimization parameters, while the constraint vector encodes information about how much the model prediction violates the conservation properties for mass, momentum and energy.

Aug-Lag algorithms have a two-level nested loop structure where the inner problem in \eqref{eq:auglag_inner} is solved using a conventional unconstrained optimization method such as the BFGS algorithm~\cite{broyden1970convergence,fletcher1970new,goldfarb1970family,shanno1970conditioning}, while the Lagrange multipliers and the penalty factor is updated in the outer loop based on the constraint violation of the inner loop solution. Although the specifics vary with implementations, all aug-Lag algorithms follow the general principle of updating the Lagrange multipliers only with sufficient improvement in the constraint violation, and increasing the penalty parameter otherwise.

This approach to solving constrained problems has largely been supplanted by constrained sequential quadratic programming (SQP) techniques~\cite{murtagh1983minos,gill2005snopt,nocedal2006knitro} that typically exhibit superior convergence properties and can produce solutions with fewer objective function and gradient evaluations than aug-Lag~\cite{bongartz1997numerical,bongartz1997complete,nocedal2006numerical}. However, optimization problems emerging from ML applications pose some challenges to SQP methods that make aug-Lag once again a relevant approach.

Recall that in Section \ref{sec:ml_overview}, we reviewed the use of stochastic mini-batch gradients due to the computational cost in evaluating loss function gradient for large training data sets. These stochastic mini-batch gradients are incompatible with constrained SQP algorithms, and due to the complexity of these methods, it is not immediately clear how they may be adapted to tolerate large stochastic inaccuracies in gradients for ML applications. The simplicity of the aug-Lag approach, on the other hand, presents some obvious avenues for adaptation that we explore.

\subsection{Inner Loop: Stochastic Gradient Descent}

For our constrained ML application, we use SGD to solve the unconstrained inner problem in \eqref{eq:auglag_inner}. This represents a stochastic generalization of the augmented Lagrangian method. However, SGD does not neatly fit into the typical convergence decisions made in the aug-Lag approach for generalized constrained problems, and the aug-Lag approach requires additional modifications to accommodate this generalization.

In a typical aug-Lag algorithm, the inner optimization problem \eqref{eq:auglag_inner} is solved to a dynamic tolerance such that $\|\nabla_p \mathcal{\hat{L}}(p_k, \lambda_k, \mu_k)\|_2 \leq \omega_k$. It is recommended that this tolerance is initially set to a very loose value and gradually tightened as the outer problem converges towards the constrained solution. Nocedal and Wright recommend initial penalty and tolerance values of $\mu_0 = 10$ and $\omega_0 = \sfrac{1}{\mu_0}$, with tolerance updates of $\omega_{k+1} = \sfrac{\omega_k}{\mu_k}$ whenever constraint violation is sufficiently reduced~\cite{nocedal2006numerical}.

Unfortunately, accurate estimates of the optimality norm, $\|\nabla_p \mathcal{\hat{L}}\|_2$, is not readily available when using stochastic mini-batch gradients. Furthermore, the $\ell_2$-norm of the gradient of the augmented Lagrangian merit function is not necessarily a meaningful termination criteria for ML applications where the primary quantity of interest is predominantly the prediction accuracy of the neural network. In the current effort, we have chosen to discard the dynamic convergence tolerance and permit the SGD algorithm to iterate through the entire training data set once per outer aug-Lag iteration. Consequently, the fixed number of SGD iterations for each inner solution is controlled by the ratio of training data set size to batch size.

Complementary to this approach, we also designate a user-defined number of aug-Lag outer iterations to evaluate for each randomized shuffling of the training data set. This means that SGD performs multiple passes over the same data set and same batches before new random batches are generated, but each pass solves a different inner optimization problem defined by updated Lagrange multipliers and penalty parameter values.

\subsection{Outer Loop: Multiplier and Penalty Updates}

Similar to dynamic convergence tolerances on the inner optimization problem, a typical aug-Lag algorithm also utilizes a dynamic constraint tolerance to make decisions on accepting Lagrange multiplier updates or increasing the penalty parameter. However, the theoretical foundation for this update is based on the assumption that the inner optimization problem is always solved to the required dynamic tolerance. Our earlier choice to solve the inner problem with SGD to a fixed number of iterations cannot provide this convergence guarantee. 

Instead, the stochastic aug-Lag algorithm accepts an update to the multipliers whenever the SGD solution to the inner problem achieves a fixed sufficient decrease in the constraint violation, $\|\mathcal{C}(p_{k+1})\|_2 \leq \eta \|\mathcal{C}_{best}\|_2$, determined by the scalar parameter $\eta$ and the previous best recorded solution for the constraints $\mathcal{C}_{best}$. The multiplier update itself, $\lambda_{k+1} = \lambda_k + \mu_k \mathcal{C}(p_{k+1})$, remains unchanged from the conventional aug-Lag method~\cite{nocedal2006numerical}. Likewise, we also use the unmodified penalty parameter increase $\mu_{k+1} = \sigma \mu_k$ with a fixed factor $\sigma$ when the SGD solution fails to satisfy the sufficient decrease criteria in the constraints.

\subsection{Adaptive Learning Rate}

In unconstrained ML applications, it is common practice to utilize learning rate schedules that gradually decay the step length for SGD iterations as the training converges towards the solution. This is somewhat analogous to line search methods for conventional optimization and nonlinear equations, with the caveat that the decaying learning rate is not intended to avoid local maxima and other undesirable stationary points. Instead, the purpose of such schedules is to improve model accuracy by preventing the training from ``circling the drain'' around the optimum solution.

The aug-Lag method uses different multiplier and penalty parameter values at each outer iteration to construct a new training subproblem with a different solution. Consequently, learning rate adaptations from the previous outer iterations can adversely affect the convergence of the new training problem. In our implementation, we utilize a schedule that decreases the learning rate when progress in the loss function has stagnated across SGD iterations. However, we have observed that this results in learning rates that are too small for the overall aug-Lag algorithm to make sufficient progress in reducing the constraint violation. To address this issue, we reset the learning rate back to its initial value for each aug-Lag iteration, which results in an independent decay rate in the learning rate for each training subproblem we construct.

\subsection{Algorithm Overview and Implementation Details}

We now provide an overview of the final stochastic aug-Lag method in Algorithm \ref{alg:stoch_auglag}. Our encoder-decoder neural network and the aug-Lag algorithm are implemented using pyTorch~\cite{paszke2019pytorch}, a popular open-source machine learning library written in Python. In addition to the building blocks for constructing neural networks, pyTorch also provides optimization algorithms and related tools such as learning rate schedules for use in solving the ML training problems. Our stochastic aug-Lag implementation is written as a wrapper around pyTorch's native SGD optimizer. For additional details on the neural network architecture, we refer the reader to our previous work using the fixed penalty method~\cite{miller2020penalty}.

\begin{rmk}
    Our stochastic augmented Lagrangian method introduces a number of user-controlled parameters. On the surface, the burden of hand-tuning these parameters may appear to be a greater challenge than tuning a single scalar parameter for fixed penalty methods. However, the dynamic updates to the Lagrange multipliers and the penalty factor based on convergence metrics during training render the training results relatively insensitive to the initial values of these parameters. In particular, our experience indicates that the aug-Lag method can train the model to high prediction accuracy (i.e. low MSE value) even when the constraint enforcement degrades as a result of poorly set penalty parameters. In order to better guide the reader in application of this method to other problems, we also share the parameter values used in our experiments in Section \ref{sec:results}.
\end{rmk}

\begin{algorithm}[H]
\SetAlgoLined
    \KwData{Update tolerance $\eta$, convergence tolerances $\epsilon_f$ and $\epsilon_c$, initial penalty $\mu_{init}$, penalty update factor $\sigma$, penalty safeguard $\mu_{\max}$, batch size $n_{batch}$, number of random shuffles $n_{shuffle}$ and number of aug-Lag iterations per shuffle $n_{aug-Lag}$}
    \KwResult{Trained model parameters $p^*$}
    $\mathcal{C}_{best} \leftarrow \mathcal{C}(p_0)$\;
    $\lambda_0 \leftarrow \mathbf{0}$\;
    \For{$\texttt{shuffle}=0,1,2,\dots,n_{shuffle}$}{
        Randomly shuffle training data set, split into batches of size $n_{batch}$ \;
        $\mu_0 = (\texttt{shuffle}+1)*\mu_{init}$\;
        \For{$k=0,1,2,\dots,n_{aug-Lag}-1$}{
            \tcp{Solve eqn.~\ref{eq:auglag_inner} with pyTorch SGD}
            $p_{k+1} \leftarrow \text{arg}\underset{p}{\min} \; \mathcal{J}(p) + \lambda_k^T \mathcal{C}(p) + \frac{\mu_k}{2}\|\mathcal{C}(p)\|_2^2$\; 
            \eIf{$\|\mathcal{C}(p_{k+1})\|_2 \leq \eta \|\mathcal{C}_{best}\|_2$}{
                \If{$f(p_{k+1}) \leq \epsilon_f$ and $\|\mathcal{C}(p_{k+1})\|_2 \leq \epsilon_c$}{
                    \tcp{Solution found}
                    Terminate with $p^* = p_{k+1}$\; 
                }
                \tcp{Multiplier update}
                $\lambda_{k+1} \leftarrow \lambda_k + \mu_k \mathcal{C}(p_{k+1})$\; 
                $\mathcal{C}_{best} \leftarrow \mathcal{C}(p_{k+1})$\;
                $\mu_{k+1} \leftarrow \mu_k$
            }{
                \tcp{Safeguarded penalty increase}
                $\lambda_{k+1} \leftarrow \lambda_k$\;
                $\mu_{k+1} \leftarrow \min(\sigma*\mu_k, \mu_{\max})$\; 
            }
        }
        $\lambda_0 \leftarrow \lambda_{n_{aug-Lag}}$\;
    }
\caption{Stochastic augmented Lagrangian method for constrained ML training.}\label{alg:stoch_auglag}
\end{algorithm}

\section{Numerical Experiments}
\label{sec:results}
We apply the stochastic augmented Lagrangian method to the constrained training problem for our encoder-decoder neural network. Our goal in the present report is to replace the fully nonlinear FPL collision operator in XGC with a computationally cheap approximation that can reproduce accurate collisional perturbations to the velocity particle distribution function for ions, $\Delta f_i$, given the current velocity particle distribution functions for both ions and electrons, $f_i$ and $f_e$, respectively.

\begin{figure}[h]
    \centering
    \includegraphics[width=0.75\textwidth]{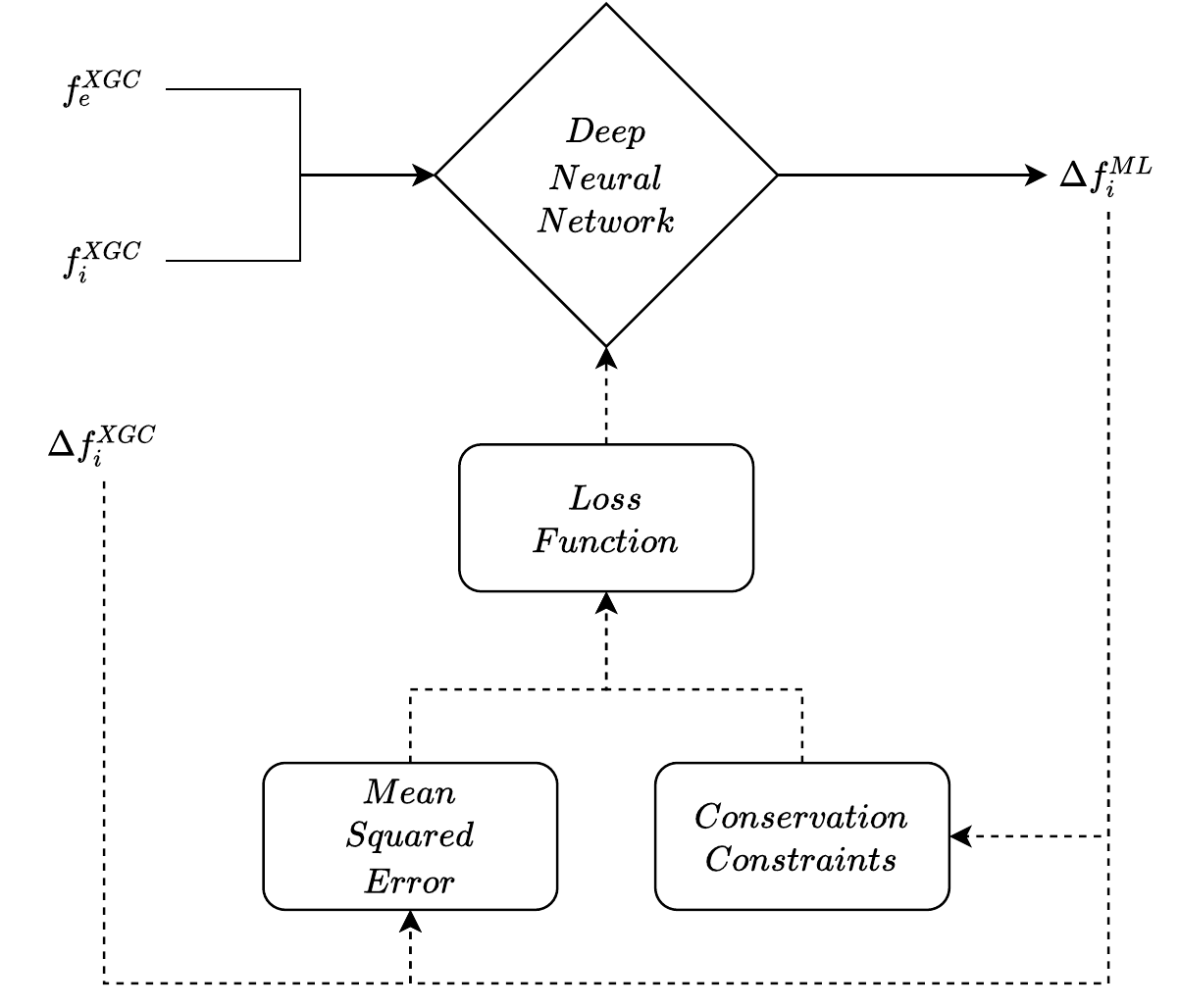}
    \caption{Flowchart for training the ML model.}\label{fig:flowchart}
\end{figure}

The training problem for this application is given as
\begin{multiline}
    \underset{p}{\text{minimize}} &\quad \mathcal{J}(p) = \frac{1}{N}\sum_j^N\|\Delta f^{ML}_{i,j}(p) - \Delta f^{XGC}_{i,j}\|_2^2 \\
    \text{subject to} &\quad \mathcal{C}(p) = \frac{1}{N}\sum_j^N\begin{pmatrix} \Delta m(\Delta f^{ML}_{i,j}(p)) \\ \Delta P(\Delta f^{ML}_{i,j}(p)) \\ \Delta E(\Delta f^{ML}_{i,j}(p)) \end{pmatrix} = \mathbf{0},
\end{multiline}
where the $ML$ superscript denotes a prediction from the ML model, $XGC$ superscript denotes the ``ground truth'' evaluation from XGC's fully nonlinear FPL collision operator, and $\mathcal{C}(p)$ represents the ML model's average violation of the conservation principles for mass $m$, momentum $P$ and energy $E$ properties. Figure \ref{fig:flowchart} provides an overview of the training process; the conservation constraints are evaluated using only the ML model output, while the XGC simulation data enters into training only through the MSE function. This training data is filtered to eliminate unconverged or inaccurate XGC solutions as detailed below.

The numerical results in this section present the conservation properties and the prediction accuracy of the constrained neural network both throughout the training process and in out-of-sample testing after training, with comparisons against both the unconstrained model and the model trained with the fixed penalty method.

\subsection{Preprocessing the Training Data}

Our raw data comes from $150,000$ evaluations of the FPL collision operator for a broad range of particle distribution functions at different combination of experimentally relevant electron/ion density, electron temperature and ion temperature. We filter this data set to eliminate unconverged or inaccurate evaluations by first discarding XGC solutions that have failed to resolve to a relative error of $10^{-6}$, and then computing the conservation properties for the remaining solutions and eliminating data points that violate conservation tolerances of $\Delta m > 10^{-10}$ for mass, $\Delta P > 10^{-7}$ for momentum, and $\Delta E > 10^{-7}$ for energy. This reduces the final data set to $129,937$ points.

\begin{figure}[h]
    \centering
    \subfloat[Raw\label{fig:xgc_raw}]{
        \includegraphics[width=0.9\textwidth,trim=4 4 4 4,clip]{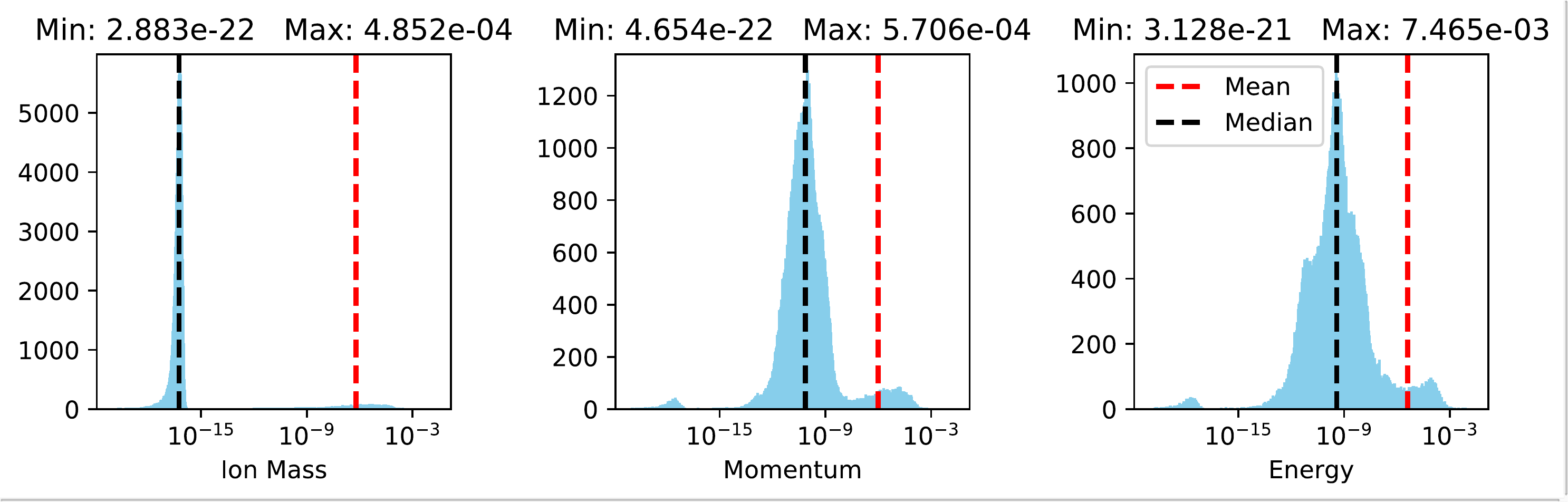}
    }
    ~\\
    \subfloat[Filtered\label{fig:xgc_cleaned}]{
        \includegraphics[width=0.9\textwidth,trim=4 4 4 4,clip]{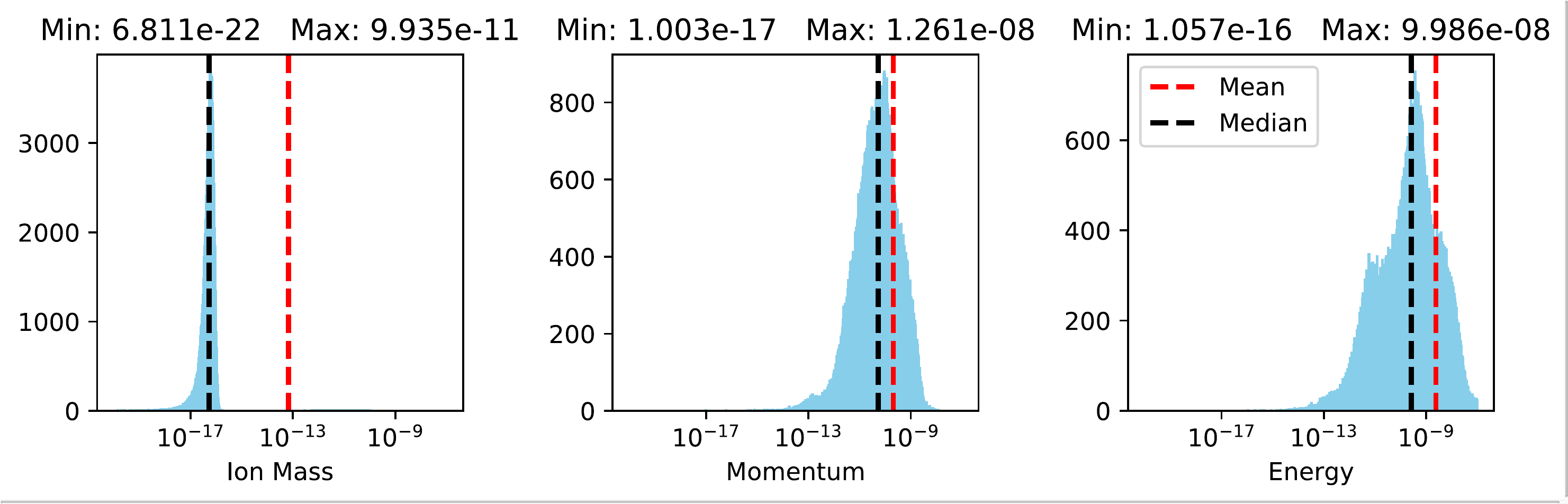}
    }
    \caption{Conservation properties for the XGC simulation data used for training, validation and testing.}\label{fig:xgc_data}
\end{figure}

Figure \ref{fig:xgc_data} below shows the histograms for the conservation properties in our raw and filtered data sets. We reserve, from random selection, $10\%$ of the filtered data set for out-of-sample tests after training. Another $10\%$ is dedicated to validating the model during training, while the remaining $80\%$ is used in solving the training problem.

\subsection{Training Results}

We train our constrained neural network using both the fixed penalty method and the proposed stochastic aug-Lag method. The hand-tuned penalty parameter for the fixed penalty method is set to $0.3$, which, for this problem, yielded the best training results by balancing the relative magnitudes of the mean squared error loss function and the $\ell_2$-norm of the conservation constraints. The aug-Lag method, on the other hand, is initialized with Lagrange multipliers set to zero, penalty factor $\mu_{init} = 100$, penalty update factor $\sigma = 2$, and penalty safeguard $\mu_{\max} = 10^9$. 

Both training methods utilize a batch size of $n_{batch} = 128$ and are limited to the same number of passes over the data set. The aug-Lag method is configured to shuffle the data $n_{shuffle} = 3$ times and perform $n_{aug-Lag} = 10$ outer aug-Lag iterations per shuffle for a total of $n_{epoch} = 30$ passes, while the fixed penalty method shuffles the data each time for all $30$ passes. The training problem is solved in single-precision on a workstation with an Intel Core i5-9400F processor, 32GB system memory, and an NVIDIA GeForce GTX 1080 GPU using CUDA version 11.0.

\begin{figure}[h]
    \centering
    \subfloat[Fixed penalty\label{fig:penalty_conv}]{
        \includegraphics[width=0.45\textwidth,trim=4 4 4 4,clip]{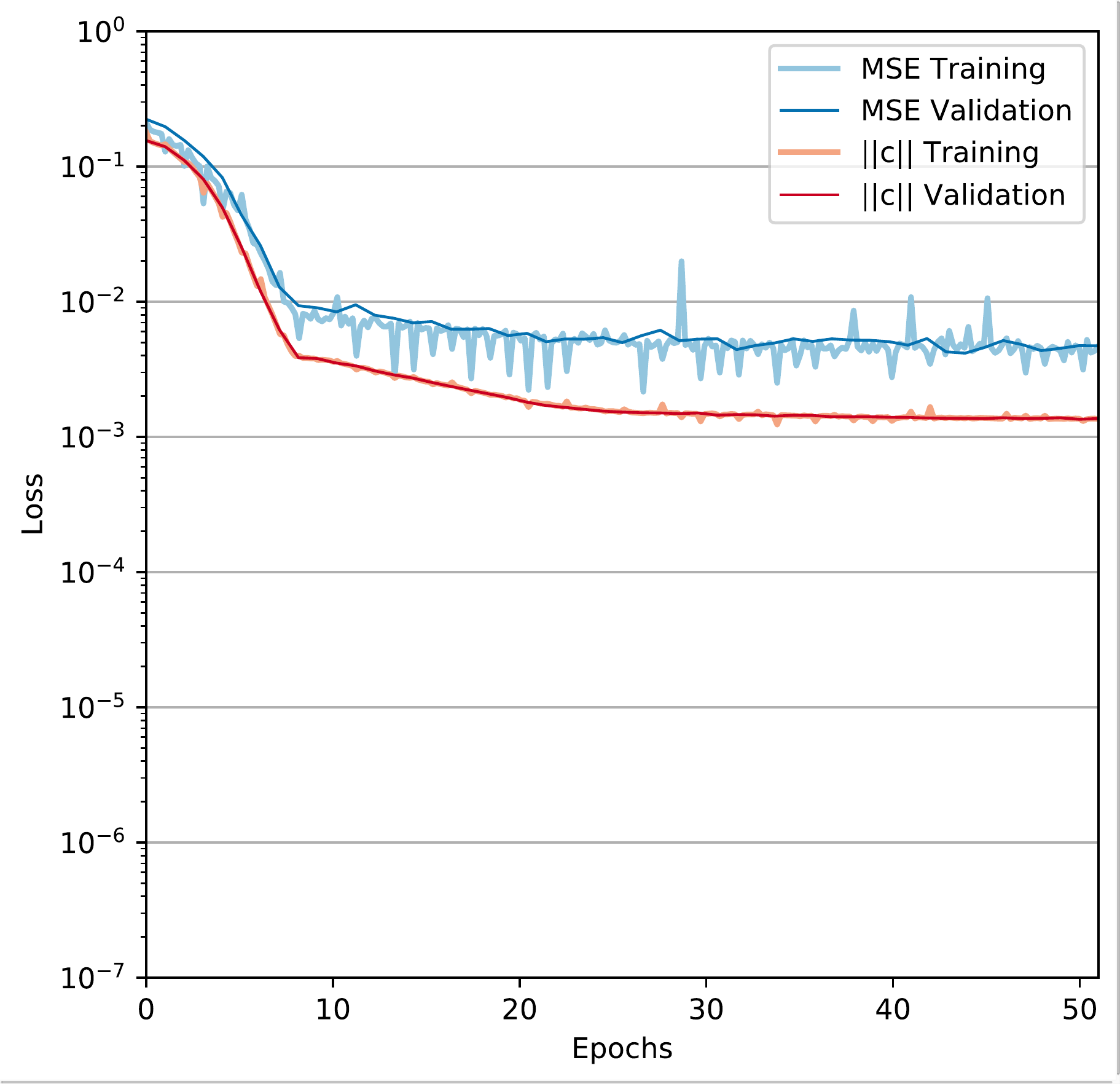}
    }
    \hfill
    \subfloat[Stochastic aug-Lag\label{fig:auglag_conv}]{
        \includegraphics[width=0.45\textwidth,trim=4 4 4 4,clip]{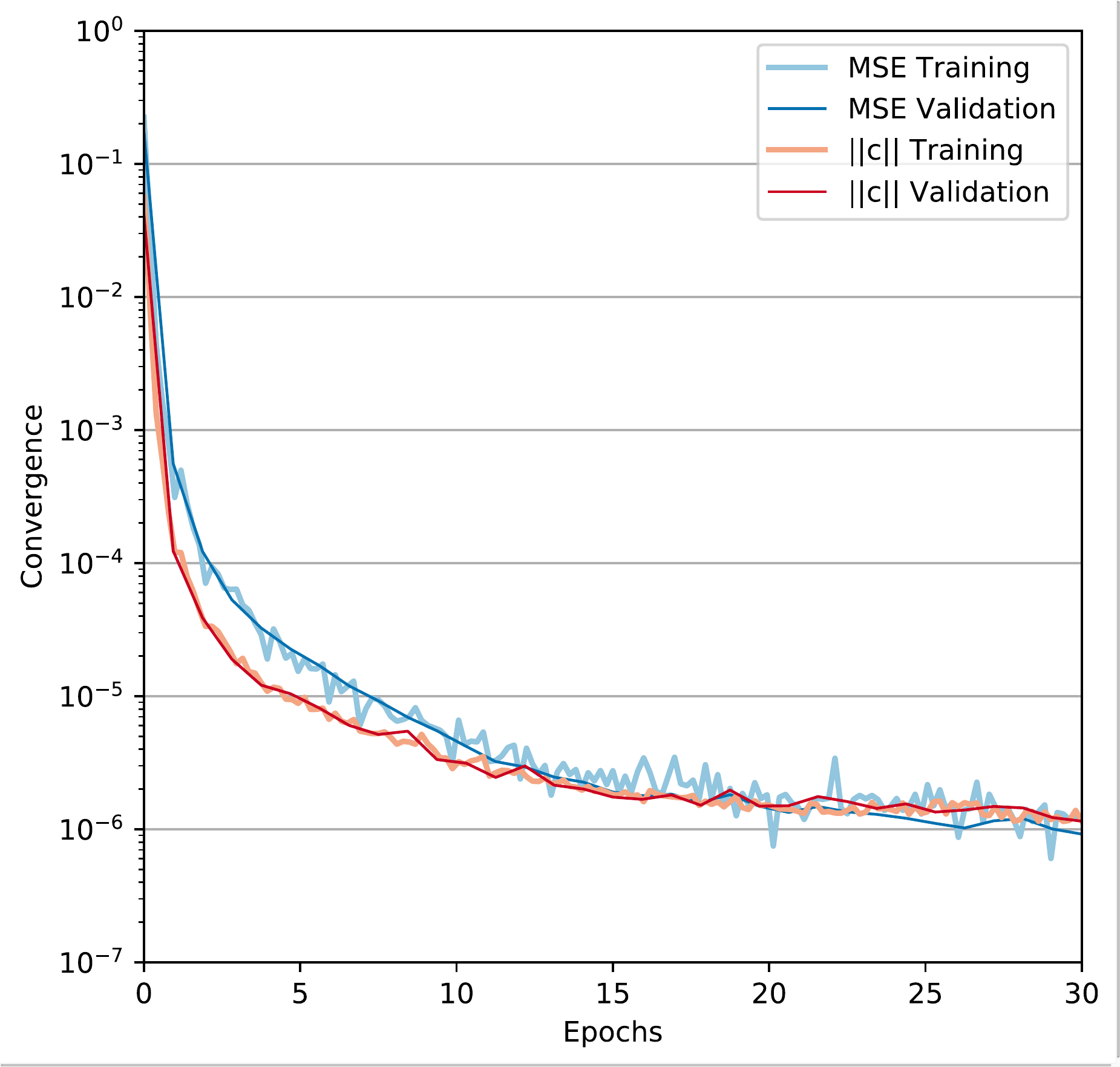}
    }
    \caption{Convergence plots for training under conservation constraints.}\label{fig:convergence}
\end{figure}

Figure \ref{fig:convergence} shows the training convergence plots tracking the mean squared error loss function value and the $\ell_2$-norm of the conservation constraints for both training methods. Our results show that the proposed stochastic aug-Lag method can train this network to a much higher accuracy using the same number of passes over the data set. An interesting observation is that the improvement in the aug-Lag method is not restricted to the enforcement of the conservation constraints. The aug-Lag method's constraint formulation clearly guides the training solution towards neural network parameters that also yield far smaller error in model predictions relative to the ground truth.

\begin{figure}[h]
    \centering
    \subfloat[Fixed penalty\label{fig:penalty_props}]{
        \includegraphics[width=0.45\textwidth,trim=4 4 4 4,clip]{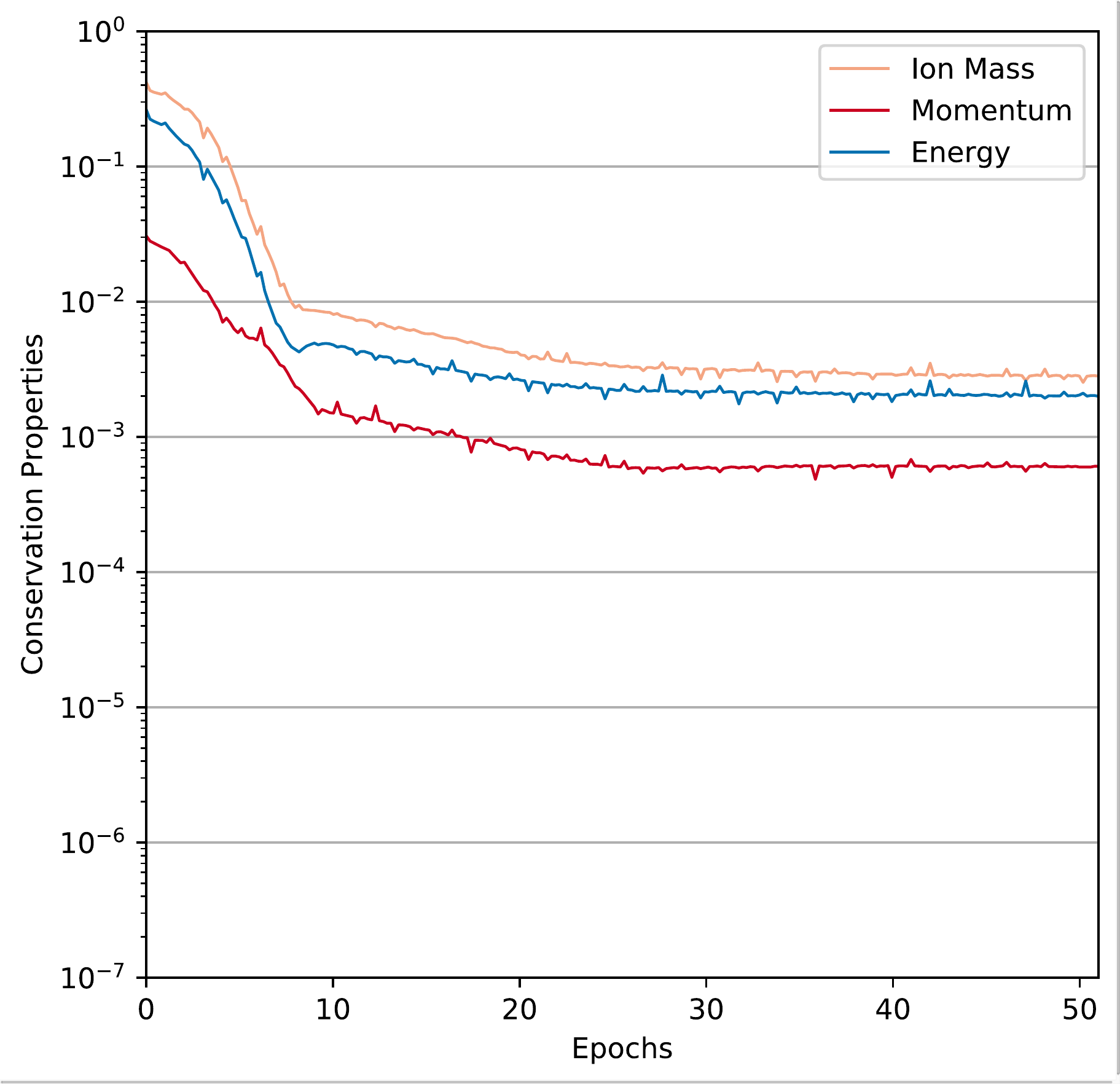}
    }
    \hfill
    \subfloat[Stochastic aug-Lag\label{fig:auglag_props}]{
        \includegraphics[width=0.45\textwidth,trim=4 4 4 4,clip]{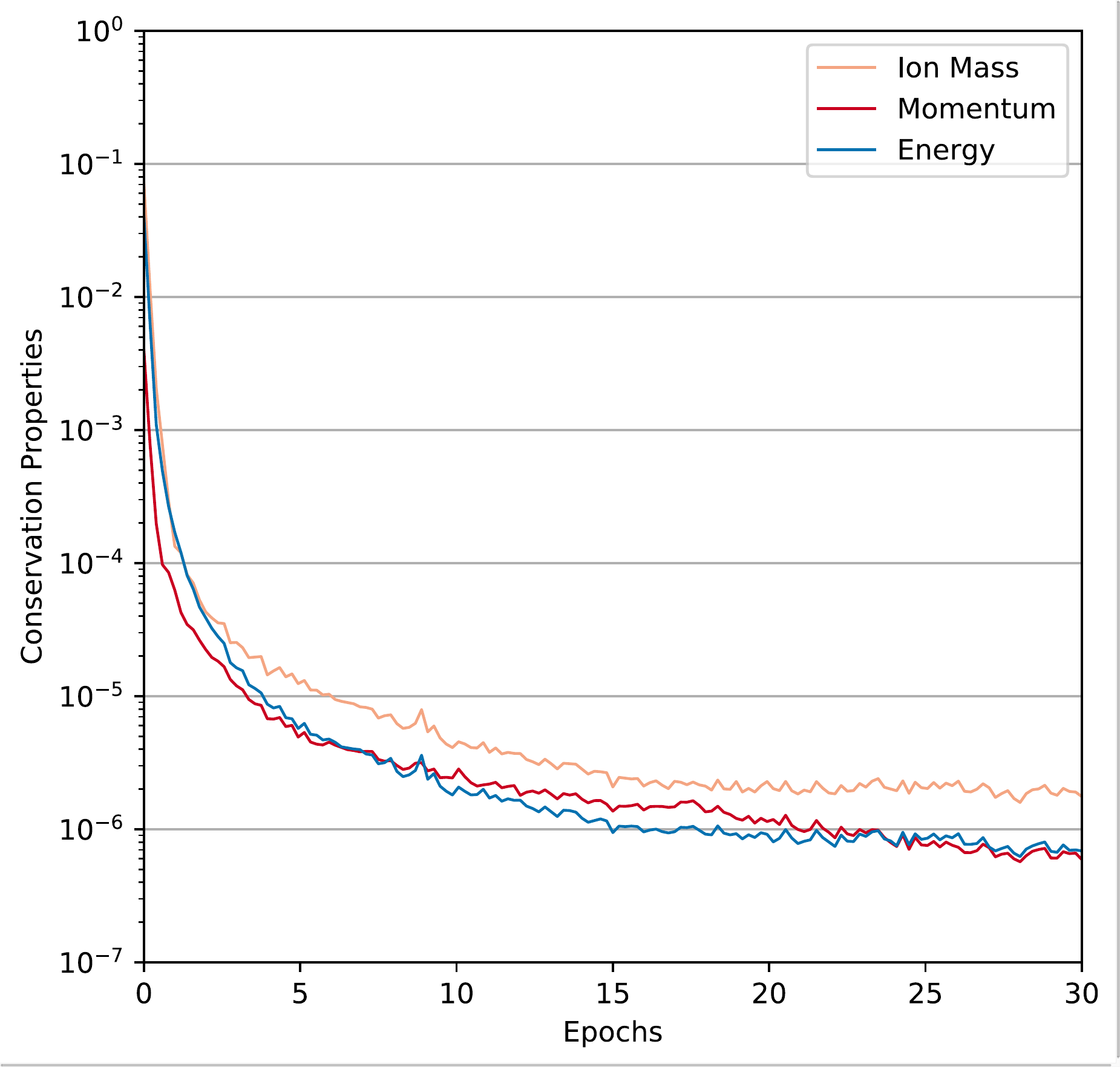}
    }
    \caption{Evolution of the convergence properties during training.}\label{fig:properties}
\end{figure}

A closer look at the individual conservation properties in Figure \ref{fig:properties} confirm that the constrained training is enforcing all three conservation properties uniformly for both methods, with the model trained with aug-Lag achieving lower constraint violation.

\begin{figure}[h]
    \centering
    \subfloat[Unconstrained\label{fig:uncon_hist}]{
        \includegraphics[width=0.9\textwidth,trim=4 4 4 4,clip]{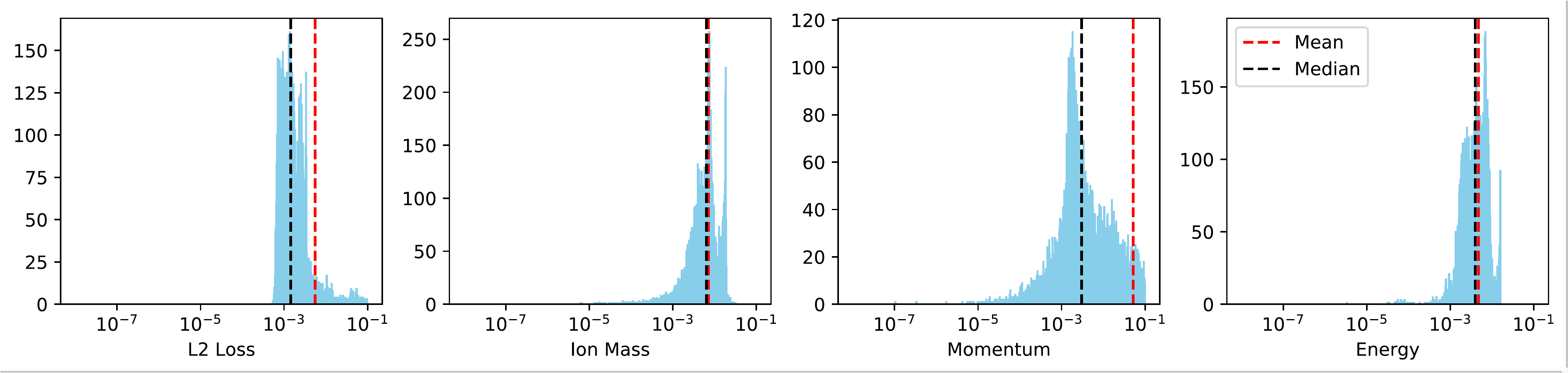}
    }
    ~\\
    \subfloat[Fixed penalty\label{fig:penalty_hist}]{
        \includegraphics[width=0.9\textwidth,trim=4 4 4 4,clip]{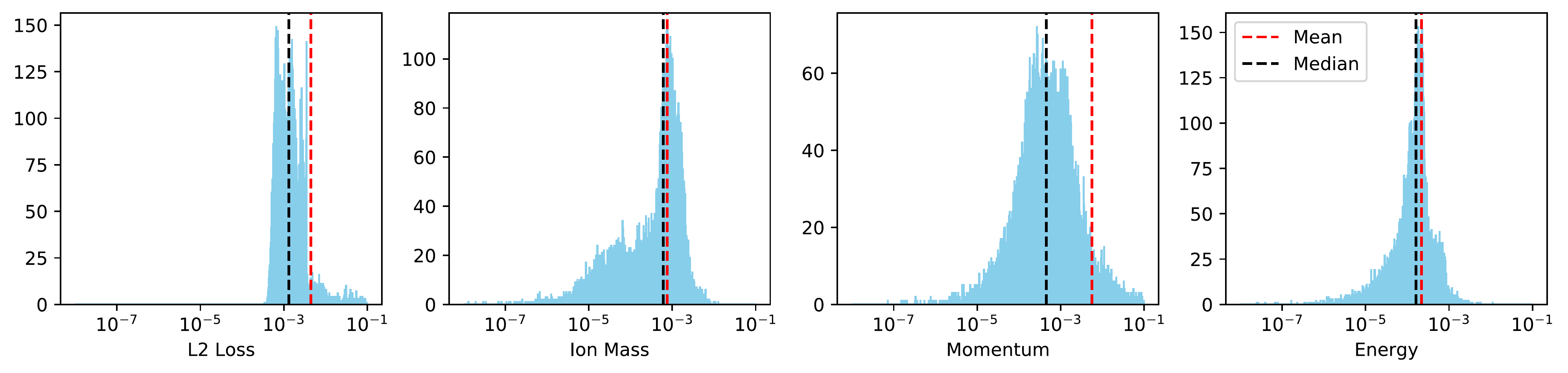}
    }
    ~\\
    \subfloat[Stochastic aug-Lag\label{fig:auglag_hist}]{
        \includegraphics[width=0.9\textwidth,trim=4 4 4 4,clip]{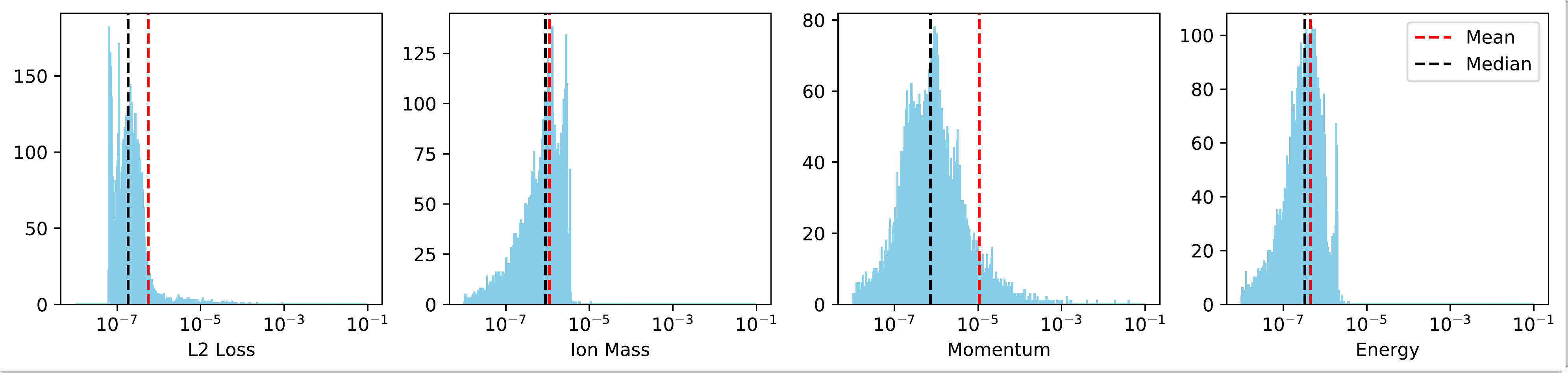}
    }
    \caption{Trained model quality in out-of-sample testing.}\label{fig:histograms}
\end{figure}

Finally, Figure \ref{fig:histograms} shows the out-of-sample tests of the constrained ML models. The unconstrained training still obeys conservation properties to a median tolerance of $10^{-3}$ simply because the underlying training data from the XGC FPL collision operator is conservative. The fixed penalty method is able to improve upon this slightly, while the aug-Lag method we have proposed in this paper is able to achieve a median constraint violation of at least $10^{-6}$ in all conservation properties that is good enough for use in XGC.

\section{Closing Remarks}
\label{sec:conclusions}
We introduced a new algorithm for solving supervised training problems in machine learning under nonlinear physical constraints on the model output. Our research is motivated by the need to replace the computationally expensive Fokker-Planck-Landau collision operator in the XGC simulation code for fusion devices. A machine learning model for efficiently approximating this collision operator must also obey the conservation laws for mass, momentum and energy properties in order to maintain the integrity of the overall XGC simulation. We construct such a model by solving a physics-constrained training problem using the proposed stochastic augmented Lagrangian method.

Our numerical results demonstrate that this training method is more effective in enforcing the conservation constraints than our previous efforts using a fixed penalty method, and achieve constraint feasibility that is sufficiently accurate for the model to be integrated into the XGC simulation. An important observation is that the tighter enforcement of the conservation constraints have guided the training towards a final model that also produces more accurate approximations of the nonlinear collision operator.

In the future, we plan to work toward integrating the trained model into the XGC workflow and to extend it to the collisional change in the electron distribution function and multiple ion speces.
Since the existing FPL solver-cost increases as square of the number of species, our ML inference time is expected to win at some cross-over number of species on exascale computers.  This study is another important topic to be addressed in the future, beginning with Summit GPUs.  We also plan to develop a more robust theoretical foundation for the heuristic multiplier and penalty parameter updates we have utilized in this effort, and investigate variations that may improve the convergence of the training problem even further.

\section*{Acknowledgements}

This work was supported by the U.S. Department of Energy, Office of Science, Office of
Advanced Scientific Computing Research, Scientific Discovery through Advanced
Computing (SciDAC) program via the FASTMath Institute under Contract
No. DE-AC02-06CH11357 at Argonne National Laboratory and via the Partnership Center for High-fidelity Boundary Plasma Simulation at Princeton Plasma Physics Laboratory under the Contract No. DE-AC02–09CH11466.

The XGC simulations used computational resources at the Argonne (Theta) and Oak Ridge (Summit)
Leadership Computing Facilities, DOE Office of Science User Facilities supported under 
Contracts DE-AC02-06CH11357 and DE-AC05-00OR22725, respectively.

\FloatBarrier

\bibliographystyle{elsarticle-num-names}
\bibliography{main.bib}

\center
\framebox{
  \parbox{4in}{
    The submitted manuscript has been created by UChicago Argonne, LLC, Operator of Argonne
    National Laboratory (``Argonne''). Argonne, operated under Contract No. DE-AC02-06CH11357, and Princeton University, Operator of Princeton Plasma Physics Laboratory under Contract No. DE-AC02-06CH11357, which are U.S. Department of Energy Office of Science
    laboratories. The U.S. Government retains
    for itself, and others acting on its behalf, a paid-up nonexclusive, irrevocable worldwide
    license in said article to reproduce, prepare derivative works, distribute copies to the
    public, and perform publicly and display publicly, by or on behalf of the Government.
    The Department of Energy will provide public access to these results of federally
    sponsored research in accordance with the DOE Public Access Plan.
    \url{http://energy.gov/downloads/doe-public-accessplan}
  }
}

\end{document}